\documentstyle[prd,twocolumn,aps]{revtex}

\begin{document}
\draft
\preprint{UCRL-JC-   }
\title{ RELATIVISTIC NUMERICAL MODEL FOR  CLOSE NEUTRON STAR BINARIES}

\author{J. R. Wilson}
\address{
University of California,
Lawrence Livermore National Laboratory,
Livermore, CA  94550}

\author{G. J. Mathews and  P. Marronetti}

\address{
University of Notre Dame,
Department of Physics,
Notre Dame, IN 46556}

\date{\today}
\maketitle
\begin{abstract}
We describe a numerical method for calculating the (3+1) dimensional general
relativistic hydrodynamics of a coalescing neutron-star binary system.  
The relativistic
field equations are solved at each time slice with a spatial 3-metric
chosen to  be conformally flat.  
Against this solution to the general relativistic field equations 
the hydrodynamic variables and gravitational radiation are allowed
to respond.  The gravitational radiation signal is
derived via a multipole expansion of the metric perturbation
to the hexadecapole ($l = 4$) order including
both mass and current moments and a correction
for the slow motion approximation.
Using this expansion, the effect of gravitational radiation 
on the system evolution can also be recovered
by introducing an acceleration term in the matter
evolution.  In the present work
we illustrate the method by applying  this model to evaluate various orbits of
two neutron stars with a gravitational mass  of 1.45 M$\odot$
near the time of the final merger.  We discuss the evidence
that, for a realistic neutron star equation of state,
general relativistic effects may cause the stars to individually
collapse into black holes prior to merging.  Also, the
strong fields cause the last stable
orbit to occur at a larger separation distance and lower frequency
than previously estimated.  
\end{abstract}
\pacs{PACS Numbers: 04.20.Jb, 04.30.+x, 47.75+f,, 95.30.Lz, 95.30.Sf
97.60.Jd, 97.80.-d}

\narrowtext

\section{INTRODUCTION}
\label{sec:level1}
Coalescing neutron stars are currently of interest for a number of reasons.
Several neutron-star binaries are known to exist in the Galaxy (e.g.~PSR 1913+16
\cite{ht75} PSR 2303+46 \cite{std85}, PSR 2127+11C \cite{agk90}, PSR 1534+11 \cite{w91})
whose orbits are observed to decay on a time scale of $1-3 \times 10^{8}$ yr.
It has been recognized for some time \cite{ce77,cvs79,t87,s86,s89,c93} that 
the final stages of coalescence of such systems could
be copious producers of gravitational radiation.  This possibility has recently
received renewed interest with the development of next generation gravity-wave
detectors such as cryogenic bar detectors \cite{bar},
 the Caltech-MIT {\it LIGO} detector \cite{a92} and its European 
counterparts, GEO and VIRGO (e.g.~ \cite{b90}) for which an event rate due 
to binary neutron star
coalescence out to 200 Mpc could be $^>_\sim 3$ per yr \cite{cvs79,nps91,p91}.  
It has also been proposed that such events (when integrated over the number of
galaxies out to high redshifts) 
could account for the observed event rate and energy requirements of
gamma-ray bursts \cite{p86,p90,p92}.  Coalescing 
neutron stars might even be significant 
contributors to
heavy element nucleosynthesis in the Galaxy \cite{ss82,m89}.

For much of the evolution of a neutron star binary, the system
should be amenable   to a  point source description using
post-Newtonian techniques\cite{lw90,kidder,wex}.  However, as the 
stars approach one another the gravitational fields become quite strong 
and hydrodynamic effects should become significant. Indeed,
it is expected that the wave forms could
become quite complex as the stars merge.
This complexity, however, may be sensitive to various physical
properties of the coalescing system \cite{c93} such as
the neutron star equation of state. Hence, careful modeling
is needed which includes both the nonlinear general relativistic effects
and a realistic neutron star equation of state.  Such calculations
can be used as a foundation for extraction of the information contained in the
detected gravity waves and as a framework in which to analyze possible 
gamma-ray burst models..

A computation of the
hydrodynamic evolution is complicated, however, due to the 
inherently three-dimensional character of the orbiting system.  To this
end several attempts have been made to model the hydrodynamics of coalescence
in either a Lagrangian smoothed-particle Newtonian approximation 
\cite{rs92,benz} or
using conventional  finite-difference methods
in the post-Newtonian approximation \cite{on89,on90,no89,ruffert}.
It important to appreciate, however, that as the two neutron stars
coalesce the system becomes strongly relativistic, and 
the validity of Newtonian or post-Newtonian hydrodynamics may be questionable.
In the present work, therefore, we improve upon such calculations in that
we explicitly include most of the effects of a fully general 
relativistic treatment.   

Some preliminary discussion of this work has  
been reported previously \cite{math1,math2,wm88,mg95,wm95,mw95}.
In this paper
we present detailed discussion of our method of solving the relativistic 
field equations and hydrodynamics. As an illustration and
for comparison with existing calculations in the literature,
we  present orbit calculations 
for two neutron stars with a gravitational mass of 1.45 M$_\odot$ each.
We find that the last stable orbit occurs for a separation distance
$\approx 1.4$ times larger and  a frequency smaller
than that estimated using the post-Newtonian
approximation.  We also find the surprising result that with a realistic equation
of state, the strong fields may induce otherwise stable neutron
stars to collapse into black holes prior to orbit instability and merging.

\section{\bf The Model}
\subsection{Coordinate System}
\label{coord}
We start with the usual slicing of spacetime into a one-parameter 
family of hypersurfaces
separated by differential displacements in 
time-like coordinates as defined in the ADM or (3+1) 
formalism \cite{adm62,mtw73}.

For this work we have considered a number of possible 
3-space coordinate systems, e.g.
polar, bipolar, spherical, cylindrical.  
Ultimately, we have selected Cartesian $x, y, z$ isotropic  coordinates.
This is a natural coordinate system
for three dimensional problems, in that no special point or singularity 
is introduced. It thus avoids problems associated with finite differencing near 
coordinate singularities.
It also has the advantage that the relativistic field equations 
assume a simpler and more symmetric form.  

With this choice for coordinates, proper distance is expressed
\begin{equation}
ds^2 = -(\alpha^2 - \beta_i\beta^i) dt^2 + 2 \beta_i dx^i dt + \gamma_{ij}dx^i dx^j
\end{equation}
where the lapse function $\alpha$ is a multiplier which 
describes the differential lapse of proper time between two
hypersurfaces.  In the Newtonian limit this quantity approaches unity
and is related to the Newtonian gravitational potential.
The quantity  $\beta_i$ is the shift vector denoting the shift in space-like
coordinates
between hypersurfaces.  The quantity $\gamma_{jk}$ is the metric of the 3-geometry.
It specifies  the distance between points within a hypersurface.

Here we introduce an approximation that the 3-geometry is both conformal and flat.
That is,  we write,
\begin{equation}
\gamma_{ij} = \phi^4  \hat \gamma_{ij} ~,
\label{conftrans}
\end{equation}
and
\begin{equation}
\hat \gamma_{ij} = \delta_{ij} ~,
\label{flat}
\end{equation}
where the conformal factor $\phi$ is a positive scalar function describing the 
ratio between the scale of distance in the curved space relative to
our flat space manifold, and $\delta_{jk}$ is the Kronecker delta.  
This is an approximate gauge condition which
we will henceforth refer to as {\it the conformally flat
condition} or {\it CFC}.
This approximation is motivated both by the general observation that
gravitational radiation in most systems studied so far is small 
\cite{ev85,ann93}, and the fact that conformal
flatness on each space-like slice considerably simplifies the solution to
the field equations.

To see the way in which the CFC allows us to solve the
relativistic field equations, consider the exact equation\cite{y79}
\begin{equation}
\dot \gamma_{ij} = -2\alpha K_{ij} + D_i \beta_j + D_j \beta_i
\label{gdot}
\end{equation}
where 
$D_i$ is the three-space covariant derivative \cite{y79}, and 
$K_{ab}$ is the extrinsic curvature describing the deformation of a
figure as it is carried forward by one unit in proper time in a direction
normal to a hypersurface.  

Equation  (\ref{gdot}) is well approximated by a conformal representation 
(\ref{conftrans}) {\it only if}
the trace free part of the right hand side vanishes.  Thus, a spatially flat
3-metric requires,
\begin{equation}
2\alpha K_{ij} = (D_i \beta_j+D_j \beta_i -{2\over 3} \gamma_{ij} D_k \beta^k)~~.
\label{detweiler}
\end{equation}
where we have also  employed the
maximal slicing condition, $tr(K_{ab}) = 0$
as a gauge choice.

We use Eq. (\ref{detweiler}) to determine the extrinsic curvature.
A convenient consequence of this  is
that any geometry which is initially conformally flat, 
will remain conformally flat to the extent that energy in gravitational
radiation is unimportant.
Equation (\ref{detweiler})
allows us to derive constraint
equations for the lapse function and conformal factor as described in the next section.

As a final condition, we take the  coordinate system 
to be rotating in such a way
as to minimize the matter
motion in the coordinate grid.  This condition
enhances the stability of the 
computation of the hydrodynamic evolution.   
However, this is a nontrivial condition to impose 
in curved spacetime which we achieve by boundary
conditions on  $\beta^i$ as described below. 
All relevant forces
are computed first in nonrotating coordinates which are then
transformed to update the the matter fields in a rotating grid.

\subsection{General Relativistic Field Equations}
For most gravitating systems studied so far (e.g.~\cite{ev85,ann93}), 
only a relatively small amount of
energy is emitted by gravitational waves.  Even for the merger
of two black holes it is expected \cite{ann93} that only a few tenths of a percent
of the rest mass will be radiated away in gravitation.
For the case of two neutron stars we would not expect any more
radiation to be emitted during the last few orbits than for a two black-hole merger,
i.e. during the inspiral, the radiated energy per orbit is a minuscule fraction
of the energy in orbital motion.
Furthermore, an explicit treatment
of the radiation reaction is exceedingly difficult \cite{ev85}.  
To treat the effects of gravitational waves
we use a multipole formalism
\cite{mtw73,Thorne}.  We use a radiation reaction potential in the
hydrodynamics equations to account for the effect of
gravity waves on the system.

The implementation of this approximation means that, 
given a distribution of mass and momentum on some manifold, we
first solve the 
constraint equations of general relativity (GR) at each time in the calculation
for a fixed distribution of matter.   
Then we let the matter and gravitational radiation respond to this geometry.
That is, we evolve
the hydrodynamic equations 
to the next time step under an assumption of ''instantaneous gravity.''
However, at each time step we obtain a time symmetric solution
to the field equations.

As an alternative to the explicit coupling 
of emitted gravitational radiation 
to the hydrodynamic and geometric evolution of the system,
the initial evolution
of the system (while the gravitational radiation
is a small perturbation) can be approximated by 
stable orbits in the absence of energy and momentum loss due to
gravitational radiation.  One can then after the fact  compute the expected
gradual loss rate of energy and momentum in gravity waves.
This latter approach is applied in some illustrative calculation 
presented here.  In a future paper we will subsequently apply
the former method to describe the late time merging and coalescence.

Our basic approach to a solution to the GR equations will be to reduce all
of the constraint equations to effective flat-space elliptic equations
which are amenable to standard techniques.  In what follows 
we make use the usual natural units in which $G = c = 1$.
Thus, at each time slice we can obtain 
a numerically valid static solution to the exact GR field equations and
information on the hydrodynamic evolution and generation of 
gravitational radiation.  However, the advance from one time 
slice to the next incorporates the approximation that the
effects of gravitational radiation can be neglected.

\subsubsection{Hamiltonian Constraint}

We begin with the Hamiltonian constraint equation \cite{y79}.
We use the forms of equations as given by Evans \cite{ev85}.
We show here that the Hamiltonian constraint and the maximal 
slicing condition ($tr(K) = 0$) can be combined so
as to form elliptic equations for both the
conformal factor $\phi$ and the product ($\alpha\phi$).

The Hamiltonian constraint equation 
can be written,
\begin{equation}
R = 16\pi \rho_H + K_{ij}K^{ij} - K^2~~,
\label{ham}
\end{equation}
where $R$ is the Ricci scalar curvature, and
\begin{equation}
\rho_H = \rho h W^2 - P~~,
\label{rhoh}
\end{equation}
where, $\rho$ is the proper baryonic matter density, $W$ is the generalization
of the special relativistic 
gamma factor ($W = \alpha U^t$, where $U^\mu$ is the four-velocity), 
$P$ is the pressure, and
$h$ is the specific relativistic enthalpy,
\begin{equation}
h = 1 + \epsilon + P/\rho~~,
\end{equation}
with $\epsilon$ the associated matter energy above the baryon rest energy,
and $P$ the matter pressure.

The conformal scaling of the 3-metric, Eq.  (\ref{conftrans}),
defines a conformal metric and manifold
($\hat \gamma,\hat M$) related to the physical metric and 
manifold ($\gamma,M$) [see refs. \cite{y79,ev85}]. 
Covariant derivatives $D_i$ and $\hat D_i$  on $M$ and $\hat M$ 
can be related by calculating the transformation of the Christoffel
connections,
\begin{equation}
\Gamma^i_{jk} = \hat \Gamma^i_{jk} + 
2 \phi^{-1}\biggl[ \delta^i_j \hat D_k \phi + 
\delta^i_k \hat D_j \phi  - \hat \gamma_{jk}\hat \gamma^{il}\hat D_l \phi \biggr]
~~.
\label{christof}
\end{equation}

With this, the transformation of the Ricci scalar curvature is
\begin{equation}
R = \phi^{-4}\hat R - 8 \phi^{-5} \hat \Delta \phi~~,
\label{ricci}
\end{equation}
where  $R = R(\gamma)$ and $\hat R = \hat R(\hat \gamma)$ and
$\hat \Delta = \hat \gamma^{ij}\hat D_i\hat D_j$.  As mentioned
in Section (\ref{coord}), we choose a conformally flat metric,
$\hat \gamma^{ij} = \delta^{ij}$, for which, $ \hat \Gamma^i_{jk}
\rightarrow 0$, $\hat D_i \rightarrow \nabla$,
$\hat R \rightarrow 0$ and $\hat \Delta 
\rightarrow \nabla^2$, the flat space Laplacian.

Solving Eq. (\ref{ricci}) for $\hat \Delta \phi$, and 
combining with the Hamiltonian constraint gives the desired form
for an elliptic equation for $\phi$,
\begin{equation}
\nabla^2 \phi = - {\phi^5 \over 8} \biggl[ 16 \pi \rho_H 
+ K_{ij} K^{ij} \biggr]~~.
\end{equation}

In order to put this constraint equation into a form which is useful
for solution along with the hydrodynamic variables, we must
introduce conformal scalings for the source terms.  To do
this, the equation of state is introduced through the adiabatic
index $\Gamma$:
\begin{equation}
P = (\Gamma - 1) \rho \epsilon~~,
\label{Gamma}
\end{equation}
where $\Gamma$ is a function of state variables for each zone.
With this equation of state, (\ref{rhoh}) becomes,
\begin{equation}
\rho_H = \rho W^2 + \rho \epsilon W \biggl[ \Gamma W -
{(\Gamma - 1) \over W} \biggr]~~. 
\label{rhoh2}
\end{equation}

For the hydrodynamic Lorentz contracted density, $D = \rho W$
and energy $E = \rho \epsilon W$, we introduce
the following conformal scalings,
\begin{equation}
D = \phi^{-6}\hat D~~,
\label{conformd}
\end{equation}
\begin{equation}
E = \phi^{-6 \Gamma}\hat E~~.
\label{conforme}
\end{equation}
The reasons for these choices will be clear when we consider
the hydrodynamic equations given in section (\ref{hydro}).

The extrinsic curvature is scaled by,
\begin{equation}
K^{ij} = \phi^{-10} \hat K^{ij}~~,
\end{equation}
which gives
\begin{equation}
K_{ij} = \phi^{-2} \hat K_{ij}~~.
\label{conformk}
\end{equation}

With the introduction of these scalings, the Hamiltonian constraint
can be written into the desired form.
\begin{eqnarray}
\nabla^2 \phi =&& - \biggl[ 2 \pi \hat D W \phi^{-1} \nonumber\\
&&+  2 \pi \hat E\biggl(\Gamma W - {\Gamma - 1 \over W}\biggr) 
\phi^{5-6\Gamma} 
 +{1 \over 8}\hat K_{ij} \hat K^{ij}\phi^{-7}\biggr]~~.
\label{phi0}
\end{eqnarray}

This can be written in a more familiar Poisson form,
\begin{equation}
\nabla^2{\phi} = -4\pi\rho_1,
\label{phi}
\end{equation}
in which the source term can be identified in terms of
physical hydrodynamic variables by transforming
the conformal scalings in Eq. (\ref{phi0}).
\begin{eqnarray}
\rho_1 =&& {\phi^5 \over 2}\biggl[DW + 
E \biggl( \Gamma W - {(\Gamma-1)\over W}\biggr)
+ {1 \over 16\pi} K_{ij}K^{ij}\biggr]~~.
\label{rho1}
\end{eqnarray}

\subsubsection{Lapse Function}

We also use the Hamiltonian constraint together with the maximal slicing condition $tr(k)=0$
 to obtain an elliptic constraint equation for the lapse function
$\alpha$.  We begin with the identities,
\begin{equation}
\Delta \alpha = \Delta [\phi^{-1}(\alpha \phi)] = D_iD^i[\phi^{-1}
(\alpha \phi)]
\end{equation}

\begin{equation}
~~~ = \phi^{-1}\Delta (\alpha \phi) -2\phi^{-6}\hat \gamma^{ij}
\hat D_i\phi \hat D_j(\alpha \phi) + \alpha \phi \Delta (\phi^{-1})~~.
\label{lapse1}
\end{equation}
Now in our conformally-flat metric one can write
for any scalar function, and in particular for the
quantity ($\alpha \phi$)
\begin{equation}
\Delta(\alpha \phi) = \phi^{-4} \hat \Delta (\alpha \phi) 
+ 2 \phi^{-5}\hat \gamma^{ij}(\hat D_i \phi)(\hat D_j(\alpha \phi))~~.
\end{equation}
Substituting this into equation (\ref{lapse1}) gives,
\begin{equation}
\Delta \alpha  = \phi^{-5} \hat \Delta(\alpha \phi) 
+ \alpha \phi \Delta (\phi^{-1})~~.
\label{alpha1}
\end{equation}
Now from the transformation properties of the Ricci 
curvature scalar (\ref{ricci}),
equation (\ref{alpha1}) can be rearranged as
\begin{equation}
\hat \Delta(\alpha \phi) = \phi^5 \Delta \alpha + {1 \over 8} \alpha 
\phi^5 \biggl[ \hat R \phi^{-4} - R \biggr]~~.
\end{equation}

Rewriting the Hamiltonian constraint (\ref{ham}) to include 
the CFC and maximal slicing conditions,
then leads to a flat-space elliptic equation in $(\alpha \phi)$.
\begin{eqnarray}
\nabla^2(\alpha \phi) =&& {1 \over 8} \alpha \phi^5\biggl[16\pi \rho
(3W^2 - 2) + \nonumber\\
&&16\pi \rho \epsilon [ 3\Gamma (W^2 + 1 ) - 5]
 + 7 K_{ij}K^{ij}\biggr]~~.
\label{alphaphi}
\end{eqnarray}
In Poisson-like  form this is
\begin{equation}
\nabla^2(\alpha\phi) = 4\pi\rho_2~~,
\label{alpha}
\end{equation}
with the source term written in terms of hydrodynamic variables as 
\begin{eqnarray}
\rho_2 = &&{\alpha \phi^5 \over 2}\biggl[{D (3W^2-2)+
E[ 3\Gamma (W^2+1)-5]\over W}\nonumber\\
&& + {7  \over 16\pi} K_{ij}K^{ij}\biggr]~~.
\label{rho2}
\end{eqnarray}
A solution of equation (\ref{alpha}) determines the lapse function after 
equation (\ref{phi}) is used to determine the conformal factor.

\subsubsection{Momentum Constraint}

With the lapse function and conformal factor determined from the 
Hamiltonian constraint and maximal slicing condition, 
we then use the momentum constraints to find
the shift vector.

The momentum constraints have the form \cite{ev85},
\begin{equation}
D_i(K^{ij} - \gamma^{ij}K) = 8 \pi S^j~~.
\label{evans37}
\end{equation}
Where $D_j$ is the three-space covariant derivative (\cite{y79}).
$S^i$ is the contravariant material momentum density
which is derived from the solution to the hydrodynamic equations,
Section (\ref{hydro}).
In our maximal-slicing conformally-flat conditions.
 the second term on the left hand
side of Eq. (\ref{evans37}) vanishes and we have,
\begin{equation}
D_i K^{ij} = 8 \pi S^j ~~.
\label{evans38}
\end{equation}
Using (\ref{conformk}) and (\ref{christof}) it can be verified
that
\begin{equation}
D_i K^{ij} = \phi^{-10}\hat D_i(\phi^{10}K^{ij})~~.
\label{evans39}
\end{equation}

Now converting our ''conformally flat condition'' [i.e.~Eq.~(\ref{detweiler})]
 from covariant 
derivatives to $\hat D^a$ or ordinary derivatives, and inserting into
equation (\ref{evans39}) gives,
\begin{equation}
D_i K^{ij} = {\phi^{-10} \over 2} \hat D_i \biggl[ {\phi^6 \over \alpha}
(\hat D^i \beta^j + \hat D^j \beta^i  - 
{2 \over 3} \hat \gamma^{ij} \hat D_k \beta^k\biggr)\biggr]~~,
\end{equation}
Combining this with (\ref{evans38})
then gives,
\begin{equation}
\hat D_i \biggl[\hat D^i \beta^j + \hat D^j \beta^i  -
 {2 \over 3} \hat \gamma^{ij} \hat D_k \beta^k)\biggr]
 = Q^j~~,
\label{wilson1}
 \end{equation}
 where the source term $Q^j$ is defined,
 \begin{equation}
 Q^j \equiv 16 \pi \alpha \phi^4 S^j + { \hat D_i \xi \over  \xi}
 \biggl[\hat D^i \beta^j + \hat D^j \beta^i  - {2 \over 3} 
 \hat \gamma^{ij} \hat D_k \beta^k \biggr]~,
\label{wilson2}
\end{equation}
 where $\xi \equiv  \alpha/\phi^6$.

Equation (\ref{wilson1}) can be reduced to,
\begin{equation}
\nabla^2 \beta^j = {\partial \over \partial x^j} \biggl({1 \over 3} 
\nabla \cdot \beta\biggr) + Q^j~.
\label{wilson3}
\end{equation}
Thus, by introducing a decomposition of $\beta^i$ into
\begin{equation}
\beta^i = B^i - {1 \over 4}\partial_i \chi~~~,
\end{equation}
the following two elliptic equations result
\begin{equation}
\nabla^2 \chi    = {4 \over 3} \nabla \cdot \beta~~~,
\label{chibeta}
\end{equation}
\begin{equation}
\nabla^2 B^i    = Q^i ~.
\label{capb}
\end{equation}
These we use to
determine the components of the shift vector.

This is the desired result except for the fact that most of the momentum
encompassed in equations (\ref{wilson2}) and (\ref{capb}) is simply the orbital motion of 
the binary.  We therefore wish to define a rotating coordinate 
system with a rotation-subtracted shift vector
in which the non-orbital aspects of matter evolution and 
relativity (e.~g.~frame drag)
can be more easily studied.  To do this we decompose $B^i$ into a
frame-drag term, $G^i$ and an
orbital motion term.
\begin{equation}
B^i = G^i + (\omega \times r)^i~~,
\label{omega}
\end{equation}
and subtract the orbital velocity of the coordinate 
system from both sides of equation (\ref{wilson3}).
Ultimately, we write equation  (\ref{capb}) in the Poisson-like form,
\begin{equation}
\nabla^2 B^i = 4 \pi \rho_3^i~~,
\label{shift}
\end{equation}
where
\begin{eqnarray}
\rho_3^i  =& & \biggl(4\alpha \phi^4 S_i - 4 \beta^iW(D + \Gamma E)\biggr)  \\
 & ~~+ & {1 \over 4\pi \xi}{\partial \xi  \over \partial x^j}
({\partial \beta^i \over \partial x^j}+{\partial \beta^j \over \partial x^i}
-{2\over 3}\delta_{ij} {\partial \beta^k\over \partial x^k})~~. \nonumber 
\end{eqnarray}
Since $\nabla^2 (\omega \times r) = 0$, $B^i$ and $G^i$ 
can be used interchangably
in equation (\ref{shift}).

As the meaning of orbital angular velocity becomes obscured in curved 
spacetime, $\omega$ in equation (\ref{omega}) takes on the meaning 
of a Lagrange multiplier which
minimizes the matter motion with respect to the coordinate system.
It only reduces to the orbital angular velocity in the Newtonian (i.e. $r 
\rightarrow \infty$) limit.
Confining  orbital motion to the $x,y$ plane, we determine
the coordinate rotation frequency
$\omega$ at each time step from the weighted average of the 
matter four velocity and the frame-drag shift vector,

\begin{equation}
\omega = {\int dV (D + \Gamma E)\biggl[{\alpha (x U_y - y U_x) \over
(1 + {U^2/ \phi^4})}\nonumber\\
 - \phi^4(x G_y - y G_x) \biggr]  \over
\int dV (D + \Gamma E) (x^2 + y^2)}
~~.
\label{domega}
\end{equation}

This rotation is then
subtracted from the velocities and added to the coordinate rotation,
thereby maintaining a centering of the stars along $y = 0$.

The fact that the constraint conditions on (\ref{phi}),
(\ref{alpha}), and (\ref{shift})
can be written in the form of flat-spaced
Poisson equations, allows for these variables to be solved by
fast numerical techniques as discussed below.  However, their solution 
requires that boundary values for these variables be specified
 at distances relatively close to the neutron stars.  Our method 
of determining the boundary values is described in section \ref{boundary}.

\subsection{Relativistic Hydrodynamics}
\label{hydro}

To solve for the fluid motions of the system in curved spacetime
it is convenient to use an Eulerian fluid description \cite{w79}.
We begin with the perfect fluid stress-energy tensor, which in covariant
form can be written,
\begin{equation}
T_{\mu\nu} = (\rho + \rho \epsilon + P)U_\mu U_\nu + P g_{\mu \nu}~~.
\end{equation}

By introducing a set of Lorentz contracted state variables it is possible to
write the relativistic hydrodynamic equations in a form which is
reminiscent of their Newtonian counterparts.  
The hydrodynamic state variables are:
The coordinate covariant baryon mass density, 
\begin{equation}
D = W \rho~~,
\end{equation}
internal energy density, 
\begin{equation}
E = W \rho \epsilon~~,
\end{equation}
three velocity,
\begin{equation}
V^i = \alpha {U_i \over \phi^4 W} - \beta^i~~,
\label{three-vel}
\end{equation}
and the momentum density,
\begin{equation}
S_i = (D + \Gamma E) U_i~~,
\label{momeq}
\end{equation}
where $W$ is a Lorentz like factor
\begin{equation}
W = \alpha U^t~ = \biggl[ 1 + {\sum{U_i^2} \over \phi^4}\biggr]^{1/2},
\end{equation}
and $\Gamma$ is an adiabatic index for the equation of
state, [Eq. (\ref{Gamma})].

In terms of these state variables, the hydrodynamic equations are as follows:
The equation for the conservation of baryon number takes the form,
\begin{equation}
{\partial D\over\partial t}  =  -6D{\partial \log\phi\over\partial t}
-{1\over\phi^6}{\partial\over\partial x^j}(\phi^6DV^j)~~.\
\end{equation}
The equation for internal energy conservation becomes,
\begin{eqnarray}
{\partial E\over\partial t}  =&&  -6\Gamma E{\partial \log\phi\over\partial t}
-{1\over\phi^6}{\partial\over\partial x^j}(\phi^6EV^j)\nonumber \\
&& -  P\biggl[{\partial W\over \partial t} +
{1\over\phi^6}{\partial\over\partial x^j}(\phi^6V^j)\biggr]~~.
\end{eqnarray}
Momentum conservation takes the form,
\begin{eqnarray}
{\partial S_i\over\partial t}& = & -6 S_i{\partial \log\phi\over\partial t}
-{1\over\phi^6}{\partial\over\partial x^j}(\phi^6S_iV^j)
-\alpha{\partial P\over \partial x^i} \\
& + & 2\alpha(D+\Gamma E)(W - {1\over W}){\partial \log\phi\over\partial
x^i} + S_j {\partial \beta^j \over \partial x^i} \nonumber \\
& - & W(D + \Gamma E){\partial \alpha \over \partial x^i} - \alpha W
(D+\Gamma E) {\partial \chi \over \partial x^i}~~, \nonumber
\label{hydromom}
\end{eqnarray}

where $\chi$ is the radiation reaction potential,
which is described in \S \ref{gw}.  Note that the repeated 
occurance of the $\phi^6$  factors simply account for proper volume factors
(proper volume $= \phi^6 (dx^i)^3$).  This
is the reason for the choice of conformal scalings introduced
in equations (\ref{conformd}) and (\ref{conforme}).  That is,
we preserve $\hat D~\hat E$, and $\hat S_i$ when $\phi$ is changed.

Our routines for evolving the 
hydrodynamics have been previously very well tested at the special
relativistic level in \cite{wm88,mcabee90,mcabee88,mcabee89,mcabee94}
where the hydrodynamics method described here was used to
simulate relativistic heavy ion calculations.  In that
work, shock-wave solutions were compared for both
decelerating and accelerating shocks.  For the former case, the
numerical results were accurate to better than 1\% over the
range of special relativistic $\gamma$-factors from 1 to 10, 
i.e. $E_{thermal} = 0~-~10~~\rho c^2$.  
For the case of
shocks accelerating matter, errors increased to $ 1$\% for
$\gamma > 2$.

A shock tube calculation was made
using a constant adiabatic index of $\Gamma = 2$.  The density
ratio was 100:1, and the initial thermal energy density
of the dense matter was equal to the initial baryonic density.
The compression ratio of the shocked bow density material
was 8\% too high.  The rest of the density profile was accurate to
better than 1\%.
Relativistic shock tube solutions were also calculated  to
test the accuracy of the rarefactions.  Again in the range of interest
($E_{thermal} \approx \rho c^2$) the overall agreement
of numerical results with exact solutions was of order 1\%.

On the basis of these results,
we anticipate all shocks occurring in the
neutron star coalescence will be treated with sufficient accuracy.
 Numerical errors are not Lorentz
invariant, but tests for invariance have shown that our numerical methods 
are reliable for changes of frame by a factor of a few in rapidity.  
In the present work we have extended the hydrodynamics to curved 
space.  However this extension is straightforward and we do not anticipate any
new instabilities or unacceptable inaccuracies to be introduced thereby.
For the calculations described here there is little fluid motion
and no strong shocks in our rotating frame.

\subsection{Equation of State}
\label{EOS}

For the orbital calculations presented here we use the zero temperature,
zero neutrino chemical potential equation of state from the 
supernova numerical model of  Mayle
 and Wilson\cite{3,5}.  
While the orbital calculations of concern in this paper should only involve
zero temperature, there is some small shock heating of the stars
as they adjust to changing conditions on the grid.
Thus, we augment this equation of state with a thermal 
component (taken to behave as a $\Gamma = 5/3$ gas)
in order to follow the dynamic evolution equations.  Thus, we write 
\begin{equation}
P = P_0(\rho) + {2 \over 3} \rho (\epsilon - \epsilon_0(\rho))~~,
\end{equation}
where $P_0$ and $\epsilon_0$ are the zero-temperature pressures and energies. 

Table 1 gives the the zero temperature values
of $P_0$, $\epsilon_0$,  and $\Gamma$ vs. $\rho$.  In figure 1
we present the baryonic mass $M_B$,
gravitational mass $M_G$, the stellar radius in Schwarzschild ($r_S$)
and isotropic ($r_I$) coordinates, the lapse function $\alpha$, and
the total mass energy density $\rho(1 +  \epsilon)$ as a
function of the central density of an isolated neutron star.  
From this figure it
can be seen that this equation of state gives an
upper limit to the gravitational mass of an isolated neutron star
of  $M_G \le 1.60$ M$_\odot$. [Note that this value supersedes
the lower value previously quoted in \cite{wm95}.]
This limit roughly agrees with the upper limit of the smallest range of
neutron star masses which overlaps all observational determinations.
The fact that this upper limit is close to the typically observed
neutron star mass $M_G \approx 1.45$  M$_\odot$ has
important consequences for the examples considered below.

The three dimensional calculations reported on here have only about
15 zones in radius to represent each neutron star.  
As another  test of the accuracy of our 3 dimensional calculations, 
therefore,
a hydrodynamic calculation was made of a single star using typical zoning in
three dimensions.  This calculation was compared
with a one dimensional spherical hydrodynamic calculation
with fine zoning.  For the same baryonic mass, 1.60 M$_\odot$, the gravitational
masses agreed to $2/3$\%, i.e. yielding a gravitational mass
of 1.45 M$_\odot$ and 1.46 M$_\odot$
for the 3D and 1D calculations respectively.
This we take as indicative of the accuracy of
the calculated gravitational binding energy of the binary system as well.  
Figure 2 shows the density profile for a single isolated $M_G = 1.45$ M$_\odot$
neutron star with the adopted equation of state..

\subsection{Gravitational Radiation}
\label{gw}

In general it is possible to express the emission of gravitational radiation
in terms of an ''exact'' expansion of multipole moments of the
effective stress energy tensor, including corrections 
for the so-called ''slow motion''
approximation\cite{Thorne}.  It is important to appreciate that these formulas apply
to strong-field sources as well as to weak field sources\cite{pt72,Thorne}
as long as the relevant components of the effective 
stress energy tensor can be identified.
Since in the present paper, we are only concerned with orbital motion
of equal mass binaries, the multipole expansions reduce to only a few nonzero
terms.  These  we evaluate and test for convergence of the expansion.
We summarize below the aspects of \cite{Thorne} which are relevant to our model. 

In any coordinate system (such as the one we are using here) in which
the gravity waves far from the source can be characterized 
as linear metric perturbations propagating on a flat background,
the transverse traceless part of the metric perturbation
characterizes the radiation completely.  This metric perturbation
can be expressed \cite{Thorne}
in terms of the mass multipole ($I^{lm}$) and
current multipole moments ($S^{lm}$) as
\begin{eqnarray}
h^{TT}_{jk} = &&\sum_{l = 2}^\infty \sum_{m = -l}^l
\biggr[ r^{-1}\ ^{(l)}I^{lm}(t-r)T^{E2,lm}_{jk}\nonumber\\
&& +  r^{-1}\ ^{(l)}S^{lm}(t-r)T^{B2,lm}_{jk}\biggr]~~,
\end{eqnarray}
where the superscript $TT$ denotes the transverse traceless
part of the metric perturbation and
the notation $^{(l)}I^{lm}$ and $^{(l)}S^{lm}$ 
denotes the $lth$ time derivative of the respective moments.

From this, the general expression for energy loss is
\begin{equation}
{dE \over dt} = {1 \over 32 \pi}\sum_{l = 2}^\infty \sum_{m = -l}^l
\langle \vert ^{(l+1)}I^{lm}\vert^2 +  \vert ^{(l+1)}S^{lm}\vert^2 \rangle ~~,
\label{dedt}
\end{equation}
where the brackets denote averages over several wavelengths.
Angular momentum loss can similarly be written
\begin{eqnarray}
{dJ \over dt} =&& {i \over 32 \pi}\sum_{l = 2}^\infty \sum_{m = -l}^l
\langle ^{(l)}I^{lm*} m ^{(l+1)}I^{lm}\rangle \nonumber\\
 &&+  \langle ^{(l)}S^{lm*} m ^{(l+1)}S^{lm}\rangle ~~~,
\label{djdt}
\end{eqnarray}
where the expression Eq. (\ref{djdt}) assumes an alignment of the
angular momentum vector with the z axis.

The radiation reaction potential $\chi$ for Eq. (\ref{momeq})
can be written
\begin{eqnarray}
\chi = {1 \over 32 \pi}\sum_{l = 2}^\infty \sum_{m = -l}^l
x_i x_j \langle \vert ^{(l+1)}I^{lm}\vert^2 +  
\vert ^{(l+1)}S^{lm}\vert^2 \rangle
  ~~~,
\label{chigrav}
\end{eqnarray}
Our problem then reduces to the identification of the relevant
mass and current moments in our coordinates.  For an asymptotically
Minkowski coordinate system one can define a quantity
\begin{equation}
\bar h^{\alpha \beta} \equiv -(-g)^{1/2} g^{\alpha \beta} + \eta^{\alpha \beta} ~~,
\end{equation}
where $g$ is the determinant of the metric and $\eta^{\alpha \beta}$ is the 
Minkowski metric tensor.  If $\bar h^{\alpha \beta}$ satisfies the de Donder gauge
condition
\begin{equation}
\bar h^{\alpha \beta} _{~,\beta} = 0 ~~,
\end{equation}
then the Einstein field equations take the form
\begin{equation}
\Box \bar h^{\alpha \beta} = -16 \pi \tau^{\alpha \beta}~~,
\label{Einstein}
\end{equation}
where $\tau^{\alpha \beta}$ is the ''effective stress-energy tensor''\cite{Thorne}.

As long as the de Donder condition is valid, Equation (\ref{Einstein})
can be inverted (using the
flat-space outgoing Green's function) and the Green's function expanded
in terms of vacuum basis functions. The resultant expression can then be
reduced \cite{Thorne} to provide expansions for the desired mass
and current moments.
\begin{eqnarray}
I^{l m}&& = {16 \pi \over (2 l + 1)!!}\biggl({(l+1)(l+2) 
\over 2 (l-1) l}\biggr)^{1/2}
\int \tau_{00} Y^{l m*} r^l d^3 x\nonumber\\
&& + \sum_{k=0}^\infty {16 \pi \over 2^k k! (2l + 2k + 1)!!} (\partial_t)^{2k}
\int \tau_{pq} r^{l+2k} \nonumber\\
&&\times \biggr[{(2l + 2k + 1) \over 2 (k+1)}\biggl({(l+1)(l+2) \over 2 (2l-1)(2l+1)}\biggr)^{1/2} T_{pq}^{2~l-2,lm*}\nonumber\\
&&+ \biggl({3(l-1)(l+2) \over (2l-1)(2l+3)}\biggr)^{1/2}  T_{pq}^{2~l,lm*}+ {2k \over 2l + 2k + 3}\nonumber\\
&& \times \biggl({l(l-1) \over 2(2l+1)(2l+3)}\biggr)^{1/2}  T_{pq}^{2~l+2,lm*}\biggr] d^3x~~,
\label{Ilm}
\end{eqnarray}
and
\begin{eqnarray}
S^{l m}&& = {-32 \pi \over (2 l + 1)!!}\biggl({(l+2)(2l+1) 
\over 2 (l-1)(l+1)}\biggr)^{1/2}\nonumber\\
&&\times \int \epsilon_{jpq} x_p(-\tau_{0q}) 
Y^{l-1,l m*}_j r^{l-1} d^3x\nonumber\\
&& + \sum_{k=0}^\infty {16 \pi i \over 2^k k! (2l + 2k + 1)!!} (\partial_t)^{2k+1}
\int \tau_{pq} r^{l+2k+1} \nonumber\\
&&\times \biggr[{1 \over 2 (k+1)}\biggl({l+2 \over 2l+1}\biggr)^{1/2} T_{pq}^{2~l-1,lm*}\nonumber\\
&&+ {1 \over 2l+2k+3}\biggl({l-1 \over 2l+1}\biggr)^{1/2}  T_{pq}^{2~l+1,lm*}\biggr] d^3x~~,
\label{Slm}
\end{eqnarray}
where the $Y^{l m*}$ are the usual spherical harmonics, and  $T_{pq}^{2~l,lm*}$ are 
the pure-orbital tensor harmonics as defined in \cite{Thorne}.  
The first integral
in Eqs. (\ref{Ilm}) and (\ref{Slm}) is the usual spherical harmonic expansion.
At the $l=2$ level, Eq. (\ref{Ilm}) reduces to the well known quadrupole approximation.
The second integral in Eqs. (\ref{Ilm}) and (\ref{Slm})
 is the correction to the slow motion
approximation, which is non negligible in the present application, 
i.e. $v/c ^>_\sim 0.1$.

To evaluate the time derivatives of the mass and current multipole moments
we make use of the rotation properties
of spherical tensors whereby, rotations can be generated
in terms of the Wigner $D$ matrices,
\begin{equation}
I^{lm} = D^l_{m m'}I_0^{lm'}~~,~~~~~~~S^{lm} = D^l_{m m'}S_0^{lm'}~~,
\label{Wignor}
\end{equation}
where $I_0^{lm'}$ and $S_0^{lm'}$ are evaluated in the rotating frame
For stable orbits (neglecting gravitational radiation) and hydrostatic stars, 
these are time independent quantities.

The main contribution to the time derivatives  is that
due to orbital motion.  Evaluation of the orbital motion
reduces to derivatives of the $ D^l_{m m'}$ which  for
our coordinates have a simple $\sim cos({m \omega t})$ dependance.

The problem with evaluating 
Equations (\ref{Ilm}) and (\ref{Slm}) is that the multipole moments are 
only defined in the de Donder gauge and not for our conformally 
flat coordinates. Furthermore, even if the transformation to our 
coordinates were straight forward (which it is not)
the effective stress energy tensor would not be known.   

Fortunately, however,
a transformation to de Donder coordinates is not necessary. 
It is only necessary that the moments of the metric
coordinates be defined in a coordinate system which, like a de Donder
coordinate system, is 
asymptotically Cartesian and mass centered (ACMC). 
In \cite{Thorne} it is proven that in such coordinate systems 
and the covariant metric components are time independent and
expandable into a spherical harmonic $(1/r)$ structure
in terms of the same moments  (i.e. Eqs. \ref{Ilm} and \ref{Slm})
relevant to the radiation field.
Furthermore, these multipole moments are invariant under transformations
between two ACMC coordinate systems.
From these expansions we can deduce the source for the slow-motion
moments to be used in the equations for the 
radiation field, (\ref{dedt}-\ref{djdt}).  For example,
the spatial three-metric must obey \cite{Thorne}:
\begin{eqnarray}
\gamma_{ij} &&= \delta_{ij} + \sum_{l=0}^N  {1 \over r^{l+1}}
\biggl[ {(2l-1)!! \over 2}
\biggl({2(l-1)l \over (l+1)(l+2)}\biggr)^{1/2}\nonumber\\
&&\times \sum_{m = -l}^{l} I^{lm} Y^{lm}
 + (l-1~pole) + \cdot\cdot\cdot + (0~pole) \biggr]~.
\label{gijt}
\end{eqnarray}
On the other hand, our spatial three metric (Eq. \ref{conftrans})
can also be expanded as the fourth power 
of a multipole expansion of the flat-space Poisson 
equation  for $\phi$ (Eq. \ref{phi}).
\begin{eqnarray}
\gamma_{ij}&& = \phi^4\delta_{ij}\nonumber\\
&& = \biggl[1 + \sum_{l = 0}^\infty \sum_{m = -l}^{l} 
{4 \pi \over (2l+1} q^{lm} Y^{lm} r^{-(l+1)}\biggr]^4\delta_{ij}~,
\label{gijw}
\end{eqnarray}
where 
\begin{equation}
q^{lm} = \int d^3x \rho_1(x) r^l Y^{lm*}~~,
\label{multipole}
\end{equation}
and $\rho_1$ is the source term for $\phi$ (Eq. \ref{rho1}).  
If we collect the dominant 
linear terms in Equations (\ref{gijt}) and (\ref{gijw})
according to the recipe given in \cite{Thorne}, then 
we can identify the relation between 
the source $\rho_1$ for the conformal factor elliptic equation
(\ref{phi}) and the mass multipole moments, i.e.
\begin{equation}
I^{lm} = {32 \pi \over (2 l + 1)!!}\biggl({(l+1)(l+2)
\over 2 (l-1) l}\biggr)^{1/2} q^{lm}~~.
\end{equation}
This identification
also reduces to the correct Newtonian limit. As
can be seen from Eq. (\ref{rho1}),
$\rho_1 \rightarrow \rho/2$, where $\rho$ is the Newtonian matter density,
so that $\tau_{00} \rightarrow \rho$ as required.

The contribution from the current moments
is expected to be small as is the slow-motion correction.  
Therefore, we are mainly
concerned with estimating the magnitude of those contributions.
To the accuracy desired, we identify the source for the current
moments $S^{lm}$ and the slow-motion corrections
with the Newtonian-like counterparts, i.e.
we set $\tau_{0j} = T_{0j}$, $\tau_{ij} = T_{ij}$.
We compute terms out to $\omega^{10}$, which includes mass multipoles
out to $l = 4$,  current multipoles out to $l = 3$ and the leading
correction for the slow motion correction.

In Table \ref{etable} we summarize the relative contributions of various
moments to the energy and angular momentum loss rates.  As expected,
the quadrupole term dominates.  The next largest term is the slow motion
correction which contributes only a few percent to the gravitational radiation
and tends to decrease the loss rate.

\subsection{Numerical Methods}

The elliptic equations for the field
and differential evolution equations
for the hydrodynamic variables
were finite-differenced in a Cartesian grid.  The intrinsic
state variables, $D, W, E, \Gamma, \alpha, \phi$,
are treated as zone centered quantities, while the four velocity $U_i$
and momentum
densities $S_i$ are node
centered.  The shift vector $\beta^i$ and
the three-velocity $V^i$ are face centered.
After finite differencing,
the elliptic equations are reduced to a matrix equation,
\begin{equation}
M \cdot x = b~~,
\end{equation}
where $M$ is a sparse matrix, $x$ is a vector representing the
relevant field variable at each zone, and $b$ is derived 
from the source terms.  This equation can then be solved
using any one of a number of fast matrix inversion techniques.

When we solve the elliptic equation for $\phi$,
the coordinate density $D$ is adjusted so as to preserve the conformal
scalings, Eqs. (\ref{conformd}) and (\ref{conforme}).  That is,
$ \hat D = \phi^6D$ is kept constant, which preserves baryon number.  Also, the coordinate 
energy density is changed to preserve $\phi^{6\Gamma}E$
and the momentum density is changed to preserve  $\phi^6 S^i$, which maintains the entropy.

\subsubsection{Extracting Physical Observables}
The gravitational mass we obtain from the asymptotic behavior
of $\phi \rightarrow 1 + (GM/2r)$ (cf Eq. \ref{gijw}).  The 
angular momentum is more
difficult to define.  We estimate this from an integral
over the space time components of the stress energy tensor \cite{mtw73}
neglecting angular momentum in the radiation field.
\begin{equation}
J^{ij} = \int \biggl(T^{i0} x^{j} - T^{j0} x^{i}\biggr) dV~~.
\end{equation}
Aligning the $z$ axis with the angular momentum vector then
gives,
\begin{equation}
J = \int (x S^y - y S^x) dV~~.
\end{equation}

\subsubsection{Boundary Conditions}
\label{boundary}

As noted above, our choices for the metric and slicing
condition lead to a form for the Hamiltonian and momentum
constraints in terms of flat-space elliptic equations, 
i.e.~ Eqs. (\ref{phi}), (\ref{alpha}), and (\ref{shift}), for the metric variables
$\phi$,  $(\alpha \phi)$ and $\beta^i$.  A solution to these
elliptic equations, however,
requires that we specify values for 
$\phi$,  $(\alpha \phi)$ and $\beta^i$ along the outer boundaries of the grid.
For a Poisson like equation, the field variables could be
specified by integrating the source function over the interior,
e.g.
\begin{equation}
\phi(x) = \int {\rho_1(x') \over \vert x - x' \vert} d^3 x'~~.
\end{equation}
However,  the evaluation of this integral for each point along the
boundaries is computationally slow.  In principle, an
expansion of the source function in spherical harmonics $Y^{lm}(\theta, \phi)$
could be applied to obtain the
field variables along the boundaries, e.g.
\begin{equation}
\phi(x) = 1 +  \sum_l^\infty \sum_{m = -l}^l
{4 \pi \over 2l+1} r^{-(l+1)}  q^{lm} Y^{lm}(\theta, \phi) ~~.
\label{spherical.harmonics}
\end{equation}
However, the convergence of
a spherical harmonic expansion for a source dominated by two
separate nearly spherical
distributions is quite slow.  In order to accommodate these two
features, we employ a combination of them which is 
numerically efficient even at distances relatively close to the neutron stars.

In the equations for $\phi$  and $(\alpha \phi)$
the  boundary values are dominated by contributions from
the effective point source potential from each star.
Our method of specifying the boundary 
for $\phi$  and $(\alpha \phi)$, therefore,  is to first make a
best fit to the source density, $\rho_1$ or $\rho_2$,
 of each star with a truncated spherical gaussian profile 
located at the source-density center of mass
in each half of the grid.  The boundary values on the
computational grid then begin with the sum of the two point-mass contributions from the
truncated spherical gaussian profiles for each star.  This provides
a simple analytic contribution around the boundary for the bulk of 
the source.

These gaussian density profiles are then subtracted from the source density to
yield a residual density.  An expansion in spherical harmonics
up to $l = 4$ (Eq.  \ref{spherical.harmonics})
is then utilized to compute the contribution from 
the residual source function over the grid.  
For stars well separated on the grid, this residual source typically
accounts for only  a few percent of the total boundary value.
Also, the spherical harmonic expansion for the residuals converges faster than
for the unsubtracted source function.  Hence, a truncation of the expansion to
$l = 4$ is sufficiently accurate.

A gaussian source profile turns out to be an excellent approximation in 
the early stages
before the neutron stars begin to coalesce.  
In future calculations in which the stars will 
be followed until they merge, however, 
they will more closely represent a single source
function.   At some point in the calculation it will become
expedient, therefore, to apply the spherical harmonic
expansion directly to the unsubtracted source function.  

We note that the expansion of the three metric (Eq. \ref{gijt})
requires that the asymptotic form for $\phi$ 
obey,
\begin{equation}
\phi \rightarrow 1 + {m_G \over 2  r}~~.
\label{limphi}
\end{equation}
Similarly, from the ACMC expansion for $g_{00}$, \cite{Thorne}
the lapse function must approach.
\begin{equation}
\alpha \rightarrow 1 - { m_G \over r}~~,
\label{limalpha}
\end{equation}
in order that our time coordinate become proper time as $r \rightarrow \infty$.
The Poisson equation  (\ref{alpha}) for ($\alpha \phi$)
can also be expanded in spherical harmonics 
(e.g. Eq. {\ref{spherical.harmonics}) yielding
\begin{equation}
(\alpha \phi) \rightarrow (\alpha \phi)_\infty - {m_{\alpha \phi} \over 2 r}~~,
\label{limalphaphi}
\end{equation}
where $m_{\alpha \phi}$ is the volume integral over twice the
source $\rho_2$.  Since in general $m_{\alpha \phi} \ne  m_G$,
we choose the boundary condition, 
\begin{equation}
(\alpha \phi)_\infty =  { m_{\alpha \phi} \over m_G}~~,
\label{bcalphaphi}
\end{equation}
to guarantee that Eq. (\ref{limalpha}) is satisfied.  For the systems 
studied here, $(\alpha \phi)_\infty \approx 0.98$.

 In our computation of the boundary conditions, we impose 
a spherical cut off in the 
matter distributions at a radius equal to the largest sphere that fits
within our cubical grid.
This avoids the possibility of a spurious hexadecapole moment associated
with the cubic grid employed in the calculation.  For matter
terms this is a reasonable truncation for the calculations presented here,
since only a negligible amount of  matter
appears beyond the surface of the neutron stars.
However, the $K_{ij}K^{ij}$ terms in  Eqs. (\ref{phi}) and  (\ref{alpha})
contribute beyond the matter boundary.  Also, the shift 
vector elliptic equations,
(\ref{chibeta}) and (\ref{capb}), involve a source which extends beyond the source boundary.

Regarding the $K_{ij}K^{ij}$ terms we note that these terms are small.
For example, the contribution to the gravitational mass from an integration
over the interior source function is only $\sim 0.0001$ M$_\odot$.  Furthermore,
the asymptotic form for  $K_{ij}K^{ij}$ should decay as $1/r^6$.  Assuming
this form, we estimate that the exterior contribution from the  $K_{ij}K^{ij}$ 
term is $_\sim^< 10^{-5}$ M$_\odot$ and can therefore be neglected
in the examples considered in this paper.

Regarding the solution for the shift vector (Eqs.  \ref{chibeta} and \ref{capb})
we note that $\nabla \cdot \beta$ is small and changes sign across the grid.
This means that the variable $\chi$ asymptotically goes to zero.
Hence, we impose  $\chi = 0$ along the boundary for Eq. \ref{chibeta}.
A solution for $B^i$ requires that we specify
the boundary condition for the ''drag'' component $G^i$.  For this
we note that $G^i$ behaves as an angular momentum density and 
should scale along the boundary  as
\begin{equation}
G^x = -{4 y  J \over r^3}~,~~  G^y = {4  x  J \over r^3}~~,
\end{equation}
\section{Orbit Calculations}
It is a nontrivial endeavor to find initial configurations 
for the two neutron stars prior to coalescence.  Our method
consists of placing two neutron stars on the grid with a rotational velocity
sufficient to keep them in orbit and an initial ''guess'' density
profile from a solution to the Tolman-Oppenheimer-Volkoff like  equation
for two single neutron stars in our isotropic coordinates.  
The conversion from single star solution to a binary solution
is achieved by allowing the stars to relax
to an equilibrium configuration on the grid.  That is,
the field equations are then solved and the hydrodynamics evolved 
(without the radiation reaction potential and with 
viscous damping of the fluid motion) until equilibrium is achieved.
For the examples to be presented below, we follow the time evolution of 
the system with constant angular momentum until it has settled down.
As the stars settle down the damping is slowly removed.

\section{Results}
\label{evolve}
In this paper we are presenting principally the method of solving
the field equations and hydrodynamics 
for binary neutron stars.  As examples, we also discuss
below three illustrative calculations made at selected values
of the orbital angular momentum with no radiation damping
of the orbits.  Highlights of these results have been presented previously
\cite{wm95}.  Here, we supply more details of the application
of the method. 

In these calculations the neutron stars are taken to be of equal mass.
The baryonic mass was selected so that in isolation each star has 
a gravitational mass of 1.45 M$_\odot$.
Although the calculations presented here ignore
radiation damping, during most of the evolution
the radiation damping is small. Therefore, the stars should follow
a sequence of quasi-equilibrium configurations which 
closely match the equilibria computed here.  These equilibria
can be analyzed to obtain the rate of energy and
momentum loss.  Ultimately, the implied orbit decay could
be used to infer the approximate time evolution through this
sequence of quasi-equilibrium orbits.

  We have placed the stars at various separation
distances on the grid and only run the calculation
long enough to check whether the
  system obtains what would be a stable orbit in the absence of gravitational
radiation.  In total for the three orbit calculations presented below
we have followed the stars through more than 20 revolutions (with roughly
two hours of CRAY/YMP  CPU time per orbit) to insure that
the orbits have had time to settle down.  We have utilized a grid
of $100\times 25 \times 25$ zones for the matter and $100 \times 50 \times 50$
for the field variables.  We make use of reflection symmetry in the orbital
plane.  Also, since here we study equal-mass binaries, we exploit reflection
inversion symmetry through the axis joining the centers of the two stars.  
In effect, then this calculation is equivalent to a 
three-space grid of 10$^6$ zones.

Initial conditions  for two
$1.60$ M$_\odot$ baryonic mass neutron stars
were obtained from the isotropic-coordinate form
of the   Tolman-Oppenheimer-Volkoff hydrostatic equilibrium.
Two single star solutions  were then 
placed on the grid and the hydrodynamic and field equations 
  evolved with viscous damping until a new quasi-equilibrium configuration
was obtained.  In practice this need only be done
once for a selected initial angular momentum.  Subsequent orbits are then
calculated by perturbing the angular momentum and allowing the system
to relax to a new stable configuration (if one exists).

The first calculation was made with an orbital angular momentum
of $2.2 \times 10^{11}$ cm$^2$.  The stars settled down into
what appeared at first as a stable orbit, but later (less than one complete orbit)
the stars began to slowly
spiral in.  For this system the angular momentum was not enough to
support the orbit.  Parameters characterizing this binary at the final
time slice calculated are given in Table \ref{paramtable}.  
[Note that this table supersedes the table in \cite{wm95}
where lower values were quoted for the central densities.]
The stars were followed to a coordinate separation $d_I \approx 34$ km
which corresponds to a ratio of proper distance to total
isolated mass of the system $m = 2 M_G^0$ of
 $d_p / m = 9.2$.  Where $M_G^0$ is the gravitational mass of a 
single isolated neutron star.
By this time it could be concluded that no stable orbit would result.  
Note that the minimum coordinate light speed $\alpha/\phi^2$
 is only 0.23  in this case.

Figure \ref{fig3} shows $x,y$ contours in the $z = 0$ plane 
for the hydrodynamic density $D$ and the
metric variables $\alpha$, and $\phi^2$.
Countours are drawn for the final time slice calculated
for this angular momentum and the other two cases studied.  

We note that even at the last time calculated 
the stars are still quite far apart
i.e. the ratio of coordinate separation distance $d_I$ to
their single-star radius, $d_I/r_I~^>_\sim 4$.  Even in
a Newtonian hydrodynamic calculation \cite{Lai1}, one would expect
at most a few percent tidal distortion at this distance.  The distortion
is made even less, however, by the strong relativistic
gravitational field (depicted in $\alpha$ and $\phi^2$ in 
Fig. \ref{fig3}) around the stars.
We also note, that the central density 
increased continuously  as the stars spiraled in.  
The orbit decay time, however is shorter than the collapse time.
Thus, it seems likely that  neither the stars nor the orbit 
are stable for this system
as summarized at the bottom of Table \ref{paramtable}.

The next calculation was made with an angular momentum of $2.3 \times 10^{11}$ 
cm$^2$.  The orbit now appeared stable as summarized at the 
bottom of Table \ref{paramtable}.
However, after about 1 to 2 revolutions the central 
densities were noticed to be rising.  By the end of the
calculation the central baryonic densities had continuously 
risen to about $2.7 \times
10^{15}$ g cm$^{-3}$ ($\approx 10$ times nuclear matter density)
which is near the maximum density for a stable
neutron star (cf. Fig. \ref{fig1}).  It appears that neutron stars of this mass
range and the adopted equation of state may continue to collapse 
as long as the released gravitational energy can be dissipated.
For this orbit the stars are at a separation distance 
of $d_p / m = 9.5$, far from merging.  However, the nonlinearities
in the gravitational field have pushed the stars over the critical density
for forming a black hole.  By the time the calculation was ended, the
minimum $\alpha$
had diminished to 0.379 and $\phi^2$ risen to 2.05
corresponding to a minimum light speed of 0.18.  Figure \ref{fig3} 
shows the stars as somewhat more compact objects which are continuing to collapse.

  The final calculation was made with the angular momentum increased
  to $2.7 \times 10^{11}$ cm$^2$.  As can be seen in Table \ref{paramtable}
and figure \ref{fig3}, the stars at this separation
$d_p / m = 12.4$ appear both stable and in a stable orbit.
Note, that even for this distance, the gravitational field
($\alpha$ and $\phi$) remains strong.

Figure \ref{fig4} shows vector fields for $U_x,U_y$, $V_x,V_y$, 
and $\beta_x,\beta_y$ for the calculation with $J = 2.3 \times 10^{11}$ cm$^2$.
Figures for the other two runs look quite similar.
The maximum value of the covariant velocity is 
$\vert U_x,U_y \vert = 0.9$.
The contravariant three velocities show the
net fluid motion after subtraction of the orbital motion.  The
maximum magnitude is $\vert V_x,V_y \vert = 0.07$.
We note that even in the corotating frame, $V_x,V_y$ exhibits some 
rotation.  Although the stars are initially corotating, 
as the stars relax to their equilibrium orbit, they acquire
net rotation relative to the orbital motion.  The net fluid motion
is still much greater than the collapse velocity which is more than
an order of magnitude slower than the maximum vector shown here. 
A peculiar aspect of the velocity field in the rotating frame
is that the fluid appears to be circulating about 
a vorticity which does not coincide with the central extrema in density
or field variables.  It also appears that the parts of the
stars most distant from the companion rotate in an opposite sense,
generating a second vorticity. 
The shift vectors are also plotted with the rotation subtracted.
What remains is frame drag around the star due to its rotation which has
a maximum amplitude of 0.015.  It is also slightly offset from 
the centers of the stars.

\subsection{Analysis of the Collapse Instability}  

  Having observed the collapse instability numerically in the neutron
star binary, one of course would like to have at least a qualitative understanding
of the source of this deviation from Newtonian intuition.  Here we present
a heuristic explanation of the observed increase in density as the stars 
approach each other. We trace this increase to the effects
of the Lorentz-like factor $W^2-1$.  This factor accounts for the
specific kinetic energy of the orbital motion of the stars.
Its effect is to increase the effective source strength.  

From Eq. (\ref{rhoh2}) the Hamiltonian density $\rho_H$  has a term 
$(W^2 - 1)(\rho + \rho \epsilon \Gamma)$ which enters into the source
term $\rho_1$ for $\phi$.  Similarly,the source for the Poisson 
equation for $(\alpha \phi)$  (cf. Eq. \ref{alphaphi})  has a term
$(W^2 - 1)\times 3 (\rho + \rho \epsilon \Gamma)$. 
Thus, the source terms for both $\phi$ and
$\alpha$ will increase as the separation distance decreases and $W$ exceeds unity.  
A stronger source term will imply larger values for both $\phi$ and
$\alpha$ at the centers of the stars and therefore steeper
gradients of these quantities as one moves outward from the center
of each star.

In isotropic coordinates, the general relativistic
condition of hydrostatic equilibrium for
each star can be inferred from the dominant terms in 
the momentum equation (51),
\begin{eqnarray}
{\partial P \over  \partial x^i} =&& -(\rho + \rho \epsilon \Gamma)
\biggl( {\partial \log{\alpha} \over \partial x^i} \nonumber \\
&& + \biggr[ {\partial \log{\alpha} \over \partial x^i} -
  2 {\partial \log{\phi} \over \partial x^i}\biggr] (W^2 - 1) \biggr)~~,
\end{eqnarray}
where  we have ignored the centrifugal term, $S_j (\partial \beta^j/\partial x^i)$.
From this we see that 
the effective gravitational force (right-hand side of Eq. 77)
increases both because $(W^2 - 1)$ exceeds unity
and because the gradients of $\alpha$ and $\phi$ are more steep
as $W^2$ increases. 
A further increase of binding arises from the $K^{ij}K_{ij}$ terms in the
field sources, but these terms are much smaller than the
$W^2-1$ contributions.

In our rotating coordinate system, the fluid three velocities $v^i$ are
nearly zero.  Hence, from Eq. (\ref{three-vel}) we have
\begin{equation}
W^2 - 1 = {1 \over (\alpha^2 /\omega^2 R^2 \phi^4) - 1}~~,
\end{equation}
where $R$ is the distance from the center of mass.  
Along the line between centers in the $J = 2.3 \times 10^{11}$ cm$^2$ model, an effective
velocity of $(\omega R \phi^2 /\alpha) = 0.28$ is obtained.
In the stars the average value is
$(W^2 - 1) \sim$ 5-10\%.  Including both the $\alpha$ and $\phi$ terms
in Eq. (77), we then estimate that
the effective hydrostatic gravitational force on the
stars is increased by 10-20\% over that of stationary non-orbiting
stars for which $(W^2 - 1) =0$.  This increased gravitational attraction
is sufficient to induce the calculated increase of central densities
as the stars approach.  An important ingredient in this analysis, however,
is that we have assumed zero thermal energy as the stars collapse, 
that is, we have assumed that neutrino emission is efficient enough
to radiate away the released gravitational binding energy so that
the stars remain effectively cold.

\subsection{Comparison with Post-Newtonian Results}

Much of the early evolution of a neutron star binary
should be describable 
with post Newtonian techniques.  However, one desires an understanding
of where in the evolution the post-Newtonian approximation 
diverges from a fully relativistic treatment.  For this reason
it is of interest to compare the results here with those obtained by
a post-Newtonian treatment.  We caution, however, that such a comparison
is ambiguous.  The two formalisms invoke different gauge choices.
Hence, parameters can have different meanings.

Our intermediate orbit ($J = 2.3 \times 10^{11}$ 
cm$^2$) appears to be on the verge of the transition from steady inspiral to
unstable plunge.  Therefore, it is convenient to compare these results
with the (post)$^{5/2}$-Newtonian analysis of this transition in 
an equal mass binary as given in ref. \cite{kidder}.  In that paper
a search was made for the inner most stable circular orbit in the
absence of radiation reaction terms in the
equations of motion.  This is analogous to the
calculations performed here which also have analyzed orbit
stability  in the
absence of radiation reaction.

In the (post)$^{5/2}$-Newtonian equations of motion, a circular orbit is
derived by setting time derivatives
of the separation, angular frequency, and the radial acceleration to zero.
This leads to the circular orbit condition \cite{kidder},
\begin{equation}
\omega_0^2 = m A_0/d_h^3~~,
\label{circular}
\end{equation}
where $\omega_0$ is the circular orbit frequency and  $m = 2M_G^0$,
$d_h$ is the
separation in harmonic coordinates, and $A_0$ is a relative 
acceleration parameter which for equal mass stars becomes,
\begin{equation}
A_0 = 1 -{3 \over 2}{m \over d_h}\biggl[3 - {77\over 8} {m \over d_h} +
(\omega_0 d_h)^2 \biggr] + {9 \over 4}(\omega_0  d_h)^2~~.
\label{a0}
\end{equation}
Equations (\ref{circular}) and (\ref{a0})
 can be solved to find the orbit angular frequency 
as a function of harmonic separation $d_h$.  
The gravity wave frequency is then twice the orbit frequency,
$f = \omega_0/\pi$.

Figure \ref{fig5} shows the circular orbit post-Newtonian
gravity wave frequency vs.
separation distance compared with the present numerical calculations.
A striking feature of the present work is that as the stars approach
one another, the frequency becomes nearly independent of the separation distance
until the orbit becomes unstable to plunge at a relatively large separation.

The main parameter characterizing the last stable orbit in the
post-Newtonian calculation is the ratio of coordinate separation to
total mass (in isolation) $d_h/m$.   
The analogous quantity in our nonperturbative
calculation is proper separation to gravitational mass, $d_P/m$.
The separation corresponding to the last stable orbit in the
post-Newtonian analysis does not occur until the stars have
approached 6.03 $m$. For $M_G^0 = 1.45 M_\odot$ stars this would correspond
to a separation distance
of about 26 km.  In the results reported here, however,
the last stable orbit occurs somewhere just below 9.4 $m_G^0$ at a proper
separation distance of $d_P \approx 40$ km.  

The fact that we observe 
the last stable orbit to occur at larger separation is consistent with the results
of Wex and Sch\"afer \cite{wex}, who found that higher-order 
 post-Newtonian terms beyond those considered in  \cite{kidder}
were significant.  They found an unstable orbit 
at 7.50 $m$ with large possible uncertainty.  
A larger separation is also consistent with the numerical
initial-data calculation for two black holes by Cook \cite{cook}.  
In that paper a minimum proper separation between {\it horizon's}
of 4.88 $m$ was found.  If the black hole
result were naively applied to the curved space around the two neutron 
stars by adding this distance to their Schwarzschild radii,
this would correspond to a separation of 8.88 $m$ between
centers.

The differences in $J/2M_G^0 \mu^0$ and angular velocity $V_\phi$
also reflect this larger distance.  Note, however,
that even though the stars are much farther apart, the relativistic
treatment gives a stronger gravitational
binding energy for this system.  
However, our binding energy includes the increased self binding of the 
individual stars.

Given that the separation distance is greater and
that $\omega \sim r^{3/2}$, we would observe 
a frequency which is a factor of $(40/26)^{3/2} \approx 2$ slower
than the post Newtonian estimates if separation distance were
the only relevant effect.  
However, our angular frequency is about a factor of two less than
the post-Newtonian value even at the same separation distance. 
We can understand this nonlinear effect of general relativity heuristically.
This difference in orbit frequency should trace
to the balance between the gravitational and centripetal forces.
From the momentum equation (51) with $W = 1$ we can write the dominant
terms as;
\begin{equation}
(D + \Gamma E) {\partial \log{\alpha} \over x^i} = S_j {\partial \beta^j 
\over x^i}~~.
\end{equation}
The left-hand side is the analog of the Newtonian gravitational force.
The right-hand side is the analog of the centripetal force.

Along the $y = 0$ axis we have  $\beta^y \approx \omega x$ and
$S_y = (D + \Gamma E) U_y$.  Thus, we may write
\begin{equation}
{\partial \log{\alpha} \over \partial x} = \omega U_y~~.
\end{equation}
Taking the fluid three velocities in the rotating frame to be zero,
$v_i = 0$, then from equation (\ref{three-vel}),
\begin{equation}
U_y = {\omega x \phi^4 \over \alpha^2 \sqrt{1 - \omega^2 \phi^4 x^2}}~~.
\end{equation}
Hence,
\begin{equation}
{\partial \log{\alpha} \over \partial x} =\omega^2 x 
{\phi^4 \over \alpha^2 \sqrt{1 - \omega^2 \phi^4 x^2}}~~.
\end{equation}
From this we see that the analog of the Newtonian centripetal term 
is enhanced by a factor of
$\phi^4/ (\alpha^2 \sqrt{1 - \omega^2 \phi^4 x^2})$.  The average
of this quantity along a line joining the centers of the stars is
$\approx 7.0$.  Thus, a much smaller value for the orbital frequency $\omega$
provides sufficient centripetal force to maintain a stable orbit.
\section{Conclusion}
We have summarized a method to study the relativistic evolution
of a binary neutron star system.  We have illustrated
the method by following the evolution of
a close binary through several orbits for different angular momenta
in the absence of radiation reaction.    These results
show two new results which to our knowledge have not been 
reported previously.  

One  significant aspect of these  calculations
is that the binary orbit becomes unstable at a much larger 
separation distance (a factor of $\approx 1.6$) than that derived
from (post)$^{5/2}$-Newtonian analysis.  
This implies that the nonlinear effects
of gravity become important much earlier on, and that searches for
gravity waves may observe the final merger to occur at a 
lower frequency than expected.  This is important, since it places
the coalescence closer to the maximum  sensitivity range of the
LIGO detector and others.  However,
we estimate little tidal distortion or hydrodynamic
amplification of the the gravity wave signal.  Our estimate
of the gravity wave amplitude near the final orbit is
$h \approx 3 \times 10^{-23}$ (at 100~Mpc), (see Table
\ref{etable} for a summary of the gravitational wave information).

A second significant result of the present calculation
is that the nonlinear effects of the
fully relativistic gravity imply deep gravitational wells
as two neutron stars approach each other.
Indeed, the fields can become so strong that
the stars are {\it individually} unstable to
collapse into two black holes.  Exactly when or if  this instability
occurs is of course dependent upon the equation of state.
However, for the realistic equation of state employed here,
this collapse is observed to occur while the stars are still
in a quasi stable orbit.  This suggests that there could be
many orbits after black-hole formation until the stars
actually merge as two black holes.
If correct, this result will have a significant impact on future studies
of binary neutron star mergers and renders the two-black-hole
coalescence much more important.  Furthermore,
the possibility of a collapse many orbit periods before
coalescence may have observational consequences not only for
gravity wave detectors, but in electromagnetic (optical,
radio, x-ray, and $\gamma$-ray) signals as well.
We note that even for our unstable inner orbit, the specific 
angular momentum $a = J/M_G^2 \approx 1.3$ Is greater than unity, implying
that more angular momentum must be lost before merger can occur,
although a transient black hole may be able to form with $a > 1$.

The importance of the possibility of premerger
stellar collapse is dependent on the equation of state and the distribution of
neutron star masses.  As stated earlier, the EOS we use is the same one 
used by Mayle \& Wilson \cite{3,5} in their supernova model.  
Calculations made with this
EOS for a model of supernova 1987A give an explosion energy of
$1.5 \times 10^{51}$ ergs, consistent with observation.
Also, the neutrino spectra and time of neutrino emission are in good
agreement with the IMB and Kamiokande neutrino detections \cite{IMB}.
These models also give a good reproduction of heavy element nucleosynthesis
in the baryon wind from the proto-neutron star \cite{Woosley94}.
An important point is that with a stiffer EOS (which would allow
a higher mass neutron star),  Mayle \& Wilson were not able to
obtain satisfactory results.

Regarding, the upper mass limit to neutron stars from observations,
Finn \cite{Finn} has recently analyzed the observed masses of neutron
stars and has assigned a lower limit of 1.15 to 1.35 M$_\odot$ and an upper limit of 
1.44 to 1.50  M$_\odot$ at the $1 \sigma$ (68\%) confidence level.  
At the $2 \sigma$ 95\%) confidence level the upper limit only increases to
1.43 to 1.64  M$_\odot$. In an independent approach,  
Bethe and Brown \cite{bb94} have recently argued from nucleosynthesis
constraints that the maximum neutron star mass is 1.56 M$\odot$.
They also point out that if kaon condensation is taken into account the critical mass
may only be 1.50 M$_\odot$.  
As seen in Fig. \ref{fig1}, our upper mass limit
is 1.60  M$_\odot$ and the mass of the stars in our sample calculations 
is 1.45  M$_\odot$, quite consistent with the observational limits.  
If the maximum observed stellar mass were as low as  the $1 \sigma$ upper limit, 
i.e.  1.50  M$_\odot$, it could be
that almost all neutron star binaries would precollapse before coalescence.

The sample calculations presented here were made with a relatively coarse 
spatial zoning 
(see \S \ref{EOS}).  We therefore, consider the present results to
be only qualitatively correct.  They must be supplemented with
more detailed numerical modeling.  We are presently 
developing methods of improved numerical
efficiency so that results will be of sufficient accuracy to allow
for a quantitative interpretation of future observed 
neutron star binary collapse and coalescence.

Note that for the above examples $\dot J/\omega J \approx 10^{-4}$ justifying 
our neglect of gravitational radiation.  Also, we point out that
since this paper was submitted it has come to our attention that
a study has been made \cite{cook2} of the validity of the conformally flat 
condition when applied to rapidly rotating isolated neutron stars.
This is the simplest case for which the conformal approximation is different
from the exact equations.  This work is encouraging in that it  has shown
that the method works remarkably well.  

 From the above discussion it is clear that further studies are
warranted, particularly an effort to better determine the last
stable orbit and the approach to this orbit, as well as a systematic
study of the sensitivity of the collapse instability
to the neutron star EOS.  Work along
this line is currently in progress.  There is also a need
to investigate orbits at larger radii so that a reliable 
connection to calculations in the post-Newtonian regime can 
be made.  Once this is done, one can combine the post-Newtonian
and numerical (3+1) results to produce a template of expected
gravity wave signals.  These can then be used 
to better extract the gravity-wave signal from the noise.

\acknowledgments

We gratefully acknowledge contributions from S. L. Detweiler
and C. R. Evans who helped us with the derivation of the field equations
and moment expansion.  We also acknowledge
contributions from Hannu Kurki-Suonio who helped us
with some of the code development. Helpful conversations with D. Eardley
and R. V. Wagoner are also gratefully acknowledged.
Work performed in part under the auspices 
of the U.~S.~Department of Energy
by the Lawrence Livermore National Laboratory under contract
W-7405-ENG-48 and NSF grant PHY-9401636.  Work at University of Notre Dame
supported in part by DOE Nuclear Theory grant DE-FG02-95ER40934.

\begin{figure}
\caption{Various parameters characterizing isolated neutron stars
with the adopted equation of state as a function of the
central baryon density $\rho_c$.  $M_B$ and $M_G$
are the baryonic and gravitational masses, respectively, in units
of M$_\odot$.  The radius is given  
in both Schwarzschild coordinates $r_S$ and
in isotropic coordinates $r_I$ in units of 10 km.  Also shown are central values for the
minimum lapse function $\alpha$,
 and the total mass energy density $\rho_{tot} = \rho(1 + \epsilon)$
(in units of $10^{15}$ g cm$^{-3}$).  }
\label{fig1}
\end{figure}

\begin{figure}
\caption{Density profile as a function of radius (in isotropic 
coordinates $r_I$) for an isolated neutron star with a gravitational
mass of 1.45 M$_\odot$.} 
\label{fig2}
\end{figure}

\begin{figure}
\caption{Contours in the orbit plane for 
various quantities at the final time slice
calculated for $J = 2.2 \times 10^{11}$ cm$^2$, $J = 2.3 \times 10^{11}$ cm$^2$,
and $J = 2.7 \times 10^{11}$ cm$^2$ as labeled.
The various figures are for: a) coordinate density $D = \rho W$; b)
lapse function $\alpha$; c) conformal factor $\phi^2$.  Note that the
$J = 2.3 \times 10^{11}$ cm$^2$ density contours are more
compact illustrating the degree to which these stars have collapsed.
See Table \ref{paramtable} for $\rho_{max}$, $\phi^2_{max}$, 
and $\alpha_{min}$. Since $W_{max} = 
1.09$, $D_{max}~^<_\sim \rho_{max}\times 1.09$.}
\label{fig3}
\end{figure}

\begin{figure}
\caption{Vectors in the $x,y$ orbit  plane around the star centered at 
$x = -17.4$ km.  This is at the final time 
calculated for the binary with $J = 2.3 \times 10^{11}$ cm$^2$.
The various figures are for: a) four velocity $U_x,U_y$ (showing 
the overall orbit motion); b)
the rotation subtracted
contravariant three-velocity $V^x , V^y$ (showing the corotation and collapse of the
fluid); and  c) the rotation
subtracted shift vectors $\beta_x , \beta_y$ (showing the relativistic 
frame drag around the star).}
\label{fig4}
\end{figure}

\begin{figure}
\caption{Comparison of gravity wave frequency $f$  vs. proper separation
distance $d_P$ from the present work (points) and post Newtonian circular orbit
condition (line).  The arrows show approximate location of the last stable
circular orbit in the two schemes.}
\label{fig5}
\end{figure}

\newpage
\baselineskip = 2\baselineskip  

\begin{table}
\caption{Zero temperature equation of state}
\begin{tabular}{cccc}
$\rho$ (g cm$^{-3})$&$\epsilon$ (erg g$^{-1})$& P (dyne cm$^{-2}$)& $\Gamma$\\
 \tableline
&&&\\
  $ 1.4656\times 10^{12}$&$1.1332\times 10^{19}$&$4.3839\times 10^{30}$&$1.2639 $\\
 $ 2.1512\times 10^{12}$&$1.2429\times 10^{19}$&$6.6872\times 10^{30}$&$1.2501 $\\
 $ 3.1575\times 10^{12}$&$1.3578\times 10^{19}$&$1.0153\times 10^{31}$&$1.2368 $\\
 $ 4.6345\times 10^{12}$&$1.4775\times 10^{19}$&$1.5364\times 10^{31}$&$1.2244 $\\
 $ 6.8024\times 10^{12}$&$1.6016\times 10^{19}$&$2.3244\times 10^{31}$&$1.2133 $\\
 $ 9.9847\times 10^{12}$&$1.7346\times 10^{19}$&$3.6993\times 10^{31}$&$1.2136 $\\
 $ 1.4656\times 10^{13}$&$1.8714\times 10^{19}$&$5.6944\times 10^{31}$&$1.2076 $\\
 $ 2.1512\times 10^{13}$&$2.0241\times 10^{19}$&$9.2761\times 10^{31}$&$1.2130 $\\
 $ 3.1575\times 10^{13}$&$3.3004\times 10^{19}$&$1.0549\times 10^{32}$&$1.1012 $\\
 $ 4.6345\times 10^{13}$&$1.2097\times 10^{19}$&$8.0239\times 10^{31}$&$1.1431 $\\
 $ 6.8024\times 10^{13}$&$1.2750\times 10^{19}$&$1.1593\times 10^{32}$&$1.1337 $\\
 $ 9.9847\times 10^{13}$&$1.3474\times 10^{19}$&$2.2347\times 10^{32}$&$1.1661 $\\
 $ 1.4656\times 10^{14}$&$1.4785\times 10^{19}$&$6.4224\times 10^{32}$&$1.2964 $\\
 $ 2.1512\times 10^{14}$&$1.7477\times 10^{19}$&$2.2614\times 10^{33}$&$1.6015 $\\
 $ 3.1575\times 10^{14}$&$2.3671\times 10^{19}$&$7.0623\times 10^{33}$&$1.9449 
$\\
 $ 4.6345\times 10^{14}$&$3.5634\times 10^{19}$&$1.8609\times 10^{34}$&$2.1268 $\\
 $ 6.8024\times 10^{14}$&$5.6410\times 10^{19}$&$4.6988\times 10^{34}$&$2.2245 $\\
 $ 9.9847\times 10^{14}$&$9.1848\times 10^{19}$&$1.1911\times 10^{35}$&$2.2988 $\\
 $ 1.4656\times 10^{15}$&$1.5307\times 10^{20}$&$3.0204\times 10^{35}$&$2.3464 $\\
 $ 2.2859\times 10^{15}$&$2.7681\times 10^{20}$&$8.9079\times 10^{35}$&$2.4078 $\\
 $ 3.1575\times 10^{15}$&$4.5386\times 10^{20}$&$2.1148\times 10^{36}$&$2.4757 $\\
 $ 4.6345\times 10^{15}$&$7.7418\times 10^{20}$&$4.6612\times 10^{36}$&$2.3016 $\\
 $ 8.1628\times 10^{15}$&$1.2080\times 10^{21}$&$1.4007\times 10^{37}$&$2.1238 $\\
 $ 9.9847\times 10^{15}$&$1.8434\times 10^{21}$&$2.0691\times 10^{37}$&$2.0865 $\\
\end{tabular}
\end{table}


\begin{table}
\caption{ Contributions to energy and momentum loss from the orbit 
calculation with $J = 2.7 \times 10^{11}$ ($cm^2$).}
\begin{tabular}{ccc}
$\dot E_{tot}$ (M$_\odot$ sec$^{-1}$) &$6.11\times 10^{-3}$\\
$\dot E_{I22}$& $6.32\times 10^{-3}$\\
$\dot E_{SM}$& $-2.2 \times 10^{-4}$\\
$\dot E_{I44}$& $2.0\times 10^{-7}$\\
$\dot E_{S32}$& $3.3\times 10^{-8 }$\\
$\dot E_{I42}$& $2.2\times 10^{-11}$\\
&&\\
$\dot J_{tot}$ (cm) &$1.07$\\
$\dot J_{I22}$& $1.11$\\
$\dot J_{SM}$& $-0.034$\\
$\dot J_{I44}$& $3.6\times 10^{-5}$\\
$\dot J_{S32}$& $-5.8\times 10^{-6}$\\
$\dot J_{I42}$& $4.2\times 10^{-9}$\\
\label{etable}
\end{tabular}
\end{table}

\begin{table}
\caption{Parameters characterizing the orbit calculations at the final 
edit. $M_G$ is just the total mass of the binary divided by 2.}
 \begin{tabular}{cccc}
 $J$ ($cm^2$)& $2.2\times 10^{11}$&$2.3\times 10^{11}$&$2.7\times 10^{11}$\\
&&\\
 \tableline
&&\\
$M_B$ (M$_\odot$)&$1.598$&$1.598$&$1.598$\\
$M_G$ (M$_\odot$)&$1.416$&$1.420$&$1.423$\\
$f$ (Hz)&$410$&$310$&$267$\\
$I^{22}$ (cm$^3$) &$1.19\times 10^{18}$&$1.28\times 10^{18}$&$2.31\times 10^{18}$\\
$d_I$ (km)&$33.8$&$34.8$&$57.0$\\
$d_P$ (km)&$39.4$&$40.6$&$53.0$\\
$\rho_{max}$ ( g cm$^{-3}$)&$2.03\times 10^{15}$&$2.70\times 10^{15}$&$1.93\times 10^{15}$\\
$W_{max}$ &$1.070$&$1.090 $&$1.085$\\
$\alpha_{min}$ &$0.440$&$0.379$&$0.463$\\
$\phi^2_{max}$ &$1.90 $&$2.05 $&$1.84 $\\
$h\cdot r$ (cm) &$1.03\times 10^4 $&$6.76 \times 10^3 $&$9.60 \times 10^3 $\\
$\dot E$ (M$_\odot$ sec$^{-1})$&$0.016$&$0.0040$& $0.0061$\\
$\dot J_{tot}$ (cm)&$1.23 $&$0.607$&$1.07 $\\
Orbit  & Unstable & Stable & Stable\\
Stars & Unstable & Unstable & Stable \\
\label{paramtable}
\end{tabular}
\end{table}

\end{document}